\documentclass[aps,pre,twocolumn,groupedaddress,showpacs,floatfix]{revtex4}
\usepackage{amsmath}
\usepackage{amssymb}
\usepackage{graphicx}

\begin{document}

\title{Scale Free Subnetworks by Design and Dynamics}

\author{Luciano da Fontoura Costa} 
\affiliation{Institute of Physics at S\~ao Carlos. 
University of S\~ ao Paulo, S\~{a}o Carlos,
SP, PO Box 369, 13560-970, 
phone +55 16 3373 9858,FAX +55 16 3371
3616, Brazil, luciano@if.sc.usp.br}

\date{27th January 2005}

\begin{abstract}   

This article addresses the degree distribution of subnetworks, namely
the number of links between the nodes in each subnetwork and the
remainder of the structure (cond-mat/0408076).  The transformation
from a subnetwork-partitioned model to a standard weighted network, as
well as its inverse, are formalized.  Such concepts are then
considered in order to obtain scale free subnetworks through design or
through a dynamics of node exchange.  While the former approach allows
the immediate derivation of scale free subnetworks, in the latter nodes
are sequentially selected with uniform probability among the
subnetworks and moved into another subnetwork with probability
proportional to the degree of the latter.  Comparison of the designed
scale-free subnetworks with random and Barab\'asi-Albert counterparts
are performed in terms of a set of hierarchical measurements.

\end{abstract}


\maketitle

\section{Introduction}

In a short period of time, complex network research progressed all the
way from uniform random
models~\cite{Flory:1941,Rapoport:1957,Erdos:1959} to the scale free
networks of Barab\'asi~\cite{Albert_Barab:2002}.  A good deal of the
motivation for such developments has been accounted for by the scale
free distribution of node degrees observed in models such as that
proposed by Barab\'asi and Albert~\cite{Albert_Barab:2002}.  One of
the principal consequences of such a type of distribution is that it
promotes the appearance of \emph{hubs}, namely nodes with particularly
high degree.  By concentrating connections, hubs play a critical role
in defining the network connectivity as well as other topological
properties such as minimal paths.  Another concept which has been
found to be particularly useful in understanding complex networks is
that of \emph{community}, which can be informally understood as a
group of nodes which are intensely interconnected but loosely
connected to the remainder of the
network(e.g.~\cite{Newman:2003,Bollt:2004,Latapy:2004,Costa_hub:2004}.

The relationship between hubs and communities has motivated some
recent works~\cite{Costa_hub:2004,Bollt:2004} which considered hubs as
references for obtaining communities.  Another concept directly
related, but not necessarily equivalent, to communities is that of a
\emph{subnetwork}~\cite{Costa_gener:2004}.  Given a network $\Gamma$,
a subnetwork of $\Gamma$ is defined as a graph including a subset of
nodes of $\Gamma$ plus their respective interconnections.  Therefore,
each community in a network can be understood as a densely linked
subnetwork which is loosely connected with the remainder of the
network.  Every community is a subnetwork, but not every subnetwork is
a community, i.e. communities are special cases of subnetworks.
Because of their generality, subnetworks represent an interesting
resource for theoretical and practical investigations of complex
networks which has only scantly been explored~\cite{Costa_gener:2004}.
One particularly interesting situation is the \emph{partition} of a
network into several subnetworks, in the sense that every node belongs
exactly to one and only subnetwork.  The concept of \emph{subnetwork
degree} was recently formalized~\cite{Costa_gener:2004} as the number
of edges linking nodes inside the subnetwork to nodes in the remainder
network.

The present work addresses subnetwork-partitioned models characterized
by scale free subnetwork degrees.  More specifically, we introduce a
transformation from scale free subnetworks to traditional weighted
networks, as well as its inverse.  Two approaches to obtain scale free
subnetworks from the random network $\Gamma$ are proposed: (i)
\emph{by design} and (ii) \emph{by dynamics}.  The former approach
starts from the desired log-log curve and applies a direct,
non-interactive method in order to obtain a subnetwork partition
having similar node degree distribution.  In the second methodology,
nodes are sequentially selected from a subnetwork and reinserted into
(possibly) another subnetwork with probability proportional to the
degree of the latter.  The comparison between the design scale free
subnetworks and traditional random and Barab\'asi-Albert models is
also considered in terms of a set of recently introduced hierarchical
features~\cite{Costa_gener:2004}.

\section{Basic Concepts} \label{sec:basic}

An undirected, \emph{unweighted} network can be represented in terms
of its \emph{adjacency} matrix $K$, such that $K(i,j)=K(j,i)=1$
whenever there is a link between nodes $i$ and $j$, with $1 \leq i,j
\leq N$, and $K(i,j)=K(j,i)=0$ otherwise.  Similarly, an undirected,
\emph{weighted} network can be represented in terms of its
\emph{weight} matrix, in the sense that $W(i,j)=W(j,i)\geq 0$
corresponds to the weight of the edge between nodes $i$ and $j$.  The
absence of edges between those nodes is represented by making
$W(i,j)=W(j,i)=0$.  Random networks, in the sense of \"Erdos and
R\'enyi~\cite{Erdos:1959,Albert_Barab:2002}, can be obtained by
selecting among the $N(N-1)/2$ possible edges with uniform probability
$\gamma$, yielding average degree $\left< k \right> = \gamma (N-1)$.

\begin{figure}
 \begin{center} 
   \includegraphics[scale=.35]{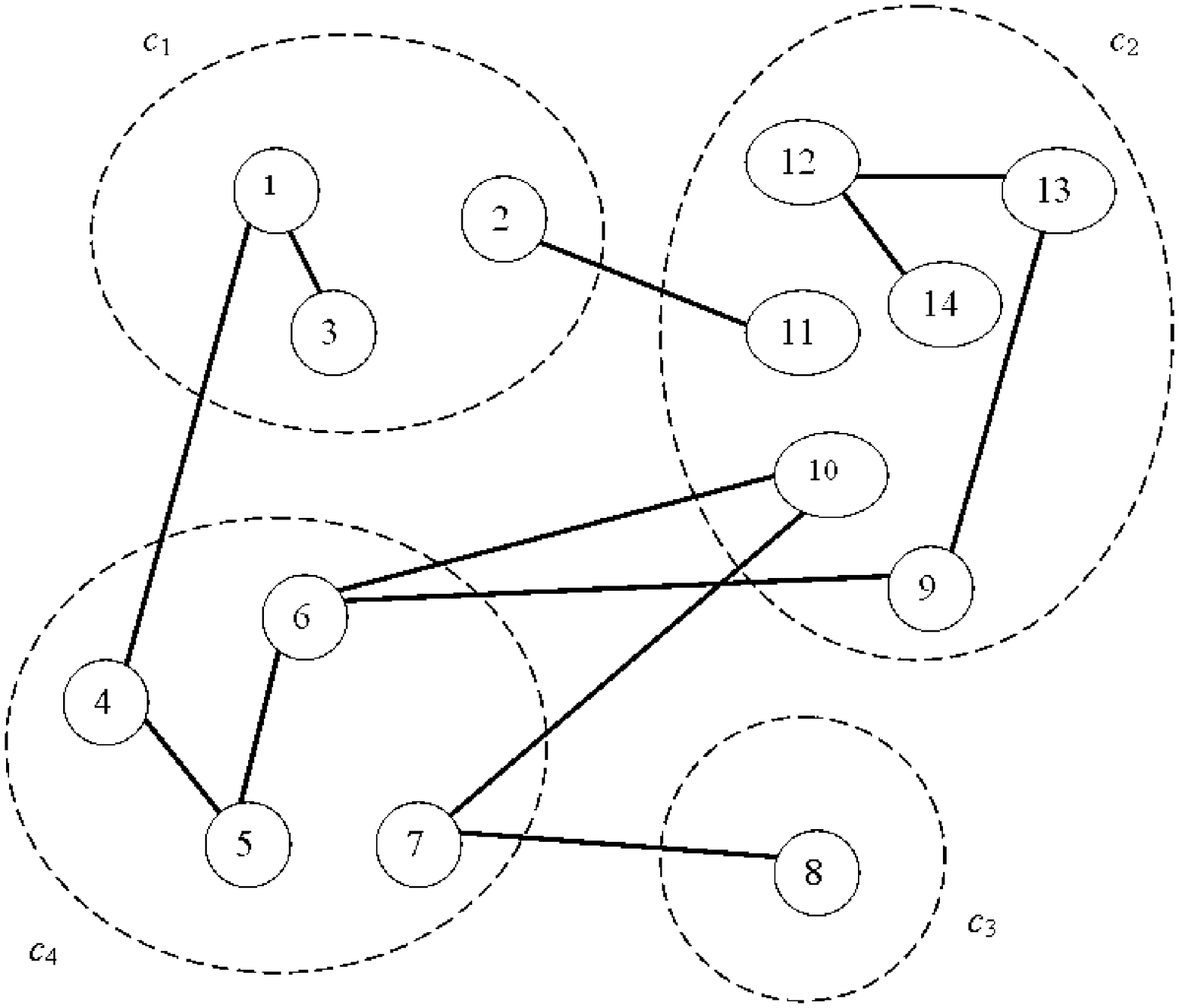}  (a) \\
   \includegraphics[scale=.3]{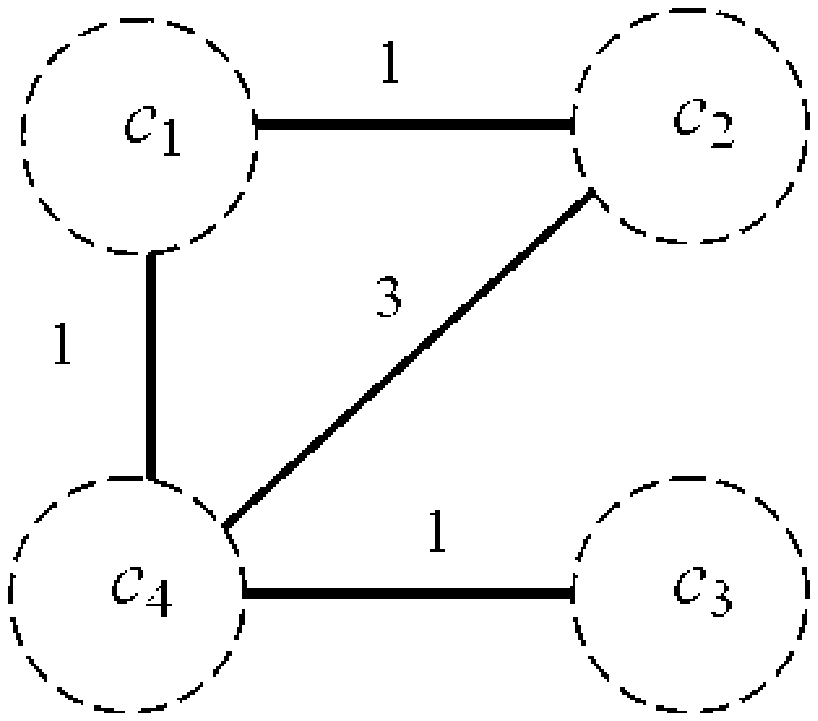}  (b)
   \caption{One of the possible subnetwork partitions of a simple
   network (a), and its respective subsumed network (b). ~\label{fig_1}}
 \end{center}
\end{figure}

The network of interest $\Gamma$ can be \emph{partitioned} into $n$
subnetworks, such that each subnetwork $c_i$ includes $N_i$ nodes from
$\Gamma$ as well as the respective interconnections.  Note that every
node should belong to exactly one subnetwork.  Figure~\ref{fig_1}(a)
illustrates a simple random network with $N=14$ nodes and its
partition into $n=4$ subnetworks.  It is henceforth assumed that the
original network $\Gamma$ to be partitioned into subnetworks follows
the \"Erdos and R\'enyi uniform model with Poisson rate $\gamma$.
Now, given any two subnetworks $c_i$ and $c_j$, the mean expected
number of edges inside each subnetwork are $e_i=\gamma N_i(N_i-1)/2$
and $e_j=\gamma N_j(N_j-1)/2$, respectively.  Let the total number of
edges in the network constituted by the two subnetworks $c_i$ and
$c_j$ be $E_{i,j} = e_i + e_j + e_{i,j}$, where $e_{i,j}$ is the
average number of edges extending between the two subnetworks.
Because $E_{i,j}=N_{i,j}(N_{i,j}-1)/2$, where $N_{i,j}=N_i+N_j$, and
$\left< k \right> = \gamma (N_{i,j}-1) \approx \gamma N_{i,j}$, it
follows that

\begin{equation}
  e_{i,j}=\gamma N_j N_i \approx \frac{N_i N_j}{N_{i,j}} \left< k \right > \label{eq:e}
\end{equation}

The degree of a subnetwork $c_i$, hence $k(c_i)$, can now be
calculated as suggested in~\cite{Costa_gener:2004}, i.e. as the number
of edges between elements of $c_i$ and the remainder of the network
$\Gamma$.  The degree of subnetwork $c_i$ can be immediately obtained
as $k(c_i) = e_{i,j}$.  Now, considering the subnetwork $c_i$ with
respect to all other $n-1$ subnetworks in the partition,
i.e. $c_j=\bigcup_{j \neq i} c_j$, we have $N_j=N-N_i$.  From
Equation~\ref{eq:e} and the fact that $\left< k \right> \approx \gamma
N$, it follows that

\begin{equation}
  k(c_i)=\gamma (N-N_i)N_i \approx \frac{N-N_i}{N} N_i \left< k \right>  \label{eq:k}
\end{equation}

In case $N \gg N_i$, we have $k(c_i) \approx N_i \left< k \right>$.

Given the original random network $\Gamma$, it is possible to
construct a subnetwork-partioned version by assigning nodes of
$\Gamma$ to each community $c_i$ according to some criterion.  The
opposite operation, namely the transformation of a partitioned network
into a traditional weighted network, henceforth called the
\emph{subsumption} of $\Gamma$ is also possible through the following
steps: (i) each community $c_i$ is subsumed into a single node $c_i$
and (ii) the weight of the edge linking two nodes $c_i$ and $c_j$ is
defined as the number of edges between the respective subnetworks.
Figure~\ref{fig_1} illustrates the subsumption of the subnetwork
partitioned structure in (a) into the weighted network in (b).  The
inverse transformation can be obtained by using the design approach
described in the following.

\section{Scale Free by Design}

In this section we present how scale free subnetwork partitions of a
random network $\Gamma$ can be immediately obtained such that the
subnetwork degree follows a pre-specified scale free distribution.

\begin{figure}
 \begin{center} 
   \includegraphics[scale=.4]{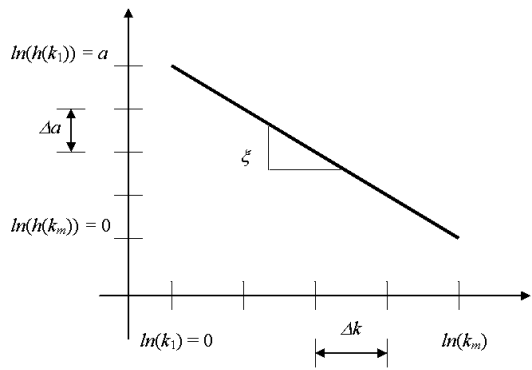}
   \caption{The basic construction used in the scale free subnetwork
   design. ~\label{fig_2}}
 \end{center}
\end{figure}

As described in the previous section, provided $N \gg N_j$, the average
degree of a subjetwork $c_j$ can be approximated as $k_j \approx N_j
\left< k \right>$, i.e. this degree becomes independent of the overall
size of the random network $\Gamma$.  This fact allows the immediate
design of subnetwork partitions following virtually any subnetwork
degree distribution, including the particularly important case of
scale free models.  The generic scale free log-log distribution of the
degrees of a network is illustrated in Figure~\ref{fig_2}.  In order
to have the subnetwork degree histogram $h(k)$ such that $h(k)
\propto k^\xi$, we start by imposing that $ln(h(k_j))=(m-j) \Delta a$
for some pre-specified $da$, with $j=1, 2, \ldots, m$, so that the
values of $ln(h(k))$ are uniformly distributed from $a$ down to 0 with
step $\Delta a = a/(m-1)$ along the $y-$axis, as $k_j$ varies from
$k_1$ to $k_m$.  It follows that $h(k_j) = exp((m-j)\Delta a)$ and
$\Delta k = - \Delta a /\xi$.  Without loss of generality, we impose
that $ln(k_1)=0$, which implies $ln(k_j)=(j-1)\Delta k$ and
$ln(k_m)=(m-1) \Delta k =-a/\xi$.  So, $ln(h(k_j))=\xi ln(k_j) +a$.

From the above developments, we have that $k_j=exp((j-1)\Delta k)$.
In other words, it is desired that community $c_j$ has degree
$k_(c_j)=k_j$.  We have from Section~\ref{sec:basic} that $k(c_j)
\approx N_j \left< k \right>$.  Therefore, in order to have $k(c_j) =
k_j$, we must have $N_j \approx k_j \left< k \right>$.  The total
required communities is $n=round(\sum_{j=1}^m h(k_j))$ and, because
$h(k_j)$ communities with $N_j$ nodes each are needed, with $j=1, 2,
\ldots m$, the total number of nodes in the random network is given as
$N=\sum_{j=1}^{m}h(k_j) N_j$.

Observe that, for a specified $h(k_j)$, the total number $N$ of nodes
can be increased by reducing $\Delta a$.

\begin{figure}
 \begin{center} 
   \includegraphics[scale=.4]{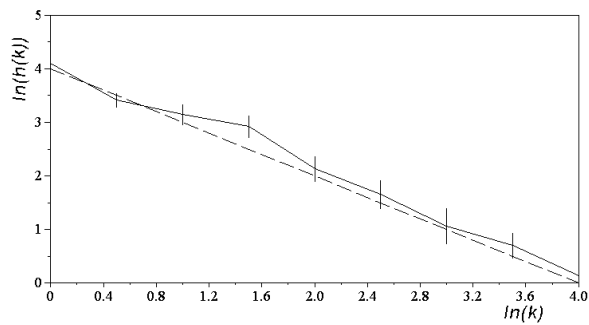}
   \caption{The average $\pm$ standard deviation of 50 realizations of
   a design scale free subnetwork partition assuming $\xi=-1.0$,
   $a=4$, $\Delta a = 0.5$ and $\left< k \right>=2$.  The dashed line
   corresponds to the originally desired distribution.~\label{fig_design}}
 \end{center}
\end{figure}

Figure~\ref{fig_design} illustrates the average $\pm$ standard
deviation of log-log node degree distributions obtained for 50
realizations of a designed subnetwork assuming $\xi=-1.0$, $a=4$,
$\Delta a = 0.5$ and $\left< k \right>=2$, implying $m=9$, $n=137$ and
$N=275$.  The obtained average curve falls reasonably close to the
desired profile (dashed straight line). The average and standard
deviation of the number of subnetworks with degree higher than zero
were 122 and 3.93, respectively.

The 50 realizations of the scale free subnetwork partitioned models
considered in the above example had their hierarchical topological
features estimated as described in~\cite{Costa_gener:2004,
Costa_hier_char:2004}.  In order to do so, the subnetwork partitions
were transformed into a traditional weighted complex network by
applying the subsumption methodology described in
Section~\ref{sec:basic}.

\begin{figure*}
 \begin{center} 
   \includegraphics[scale=1.3]{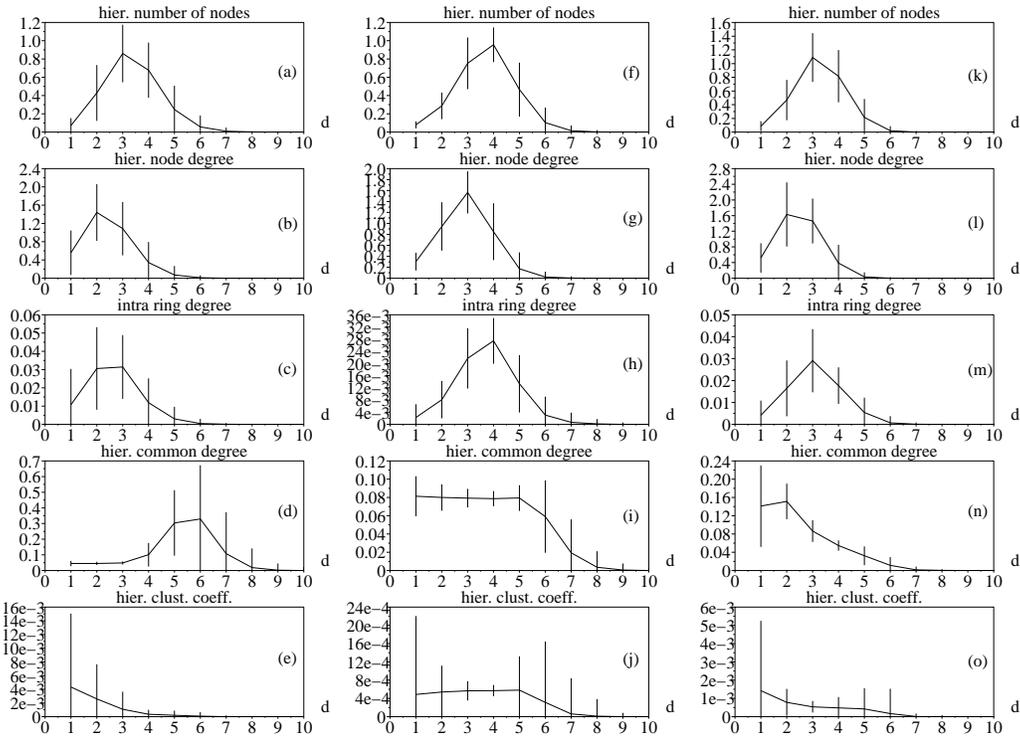}
   \caption{The average $\pm$ standard deviation of the 5 hierarchical
   measurements in terms of $d$ considering the
   50 design simulations (a-e) and random (f-j) and Barab\'asi-Albert
   (k-o) models with the same number of nodes and average
   degree. ~\label{fig_hier}}
 \end{center}
\end{figure*}

Let $R_d(i)$ be the ring defined as the subnetwork including the nodes
at minimal distance $d$, corresponding to the hierarchical level, from
a reference node $i$ and the edges between such nodes.  The considered
measurements include the average (over all nodes) of: (i) the
\emph{hierarchical number of nodes}, i.e. the number of nodes in
$R_d(i)$; (ii) the \emph{hierarchical node degree}, defined as the
number of edges between rings $R_d(i)$ and $R_{d}(i+1)$; (iii) the
\emph{intra ring degree}, i.e. the average degree among the elements
of $R_d(i)$; (v) the \emph{common degree}, namely the average of the
traditional node degree considering the nodes in $R_d(i)$; and (vi)
the \emph{hierarchical clustering coefficient}, corresponding to the
clustering coefficient of $R_d(i)$.  Figure~\ref{fig_hier} presents
the average $\pm$ standard deviations of such measurements obtained
for the above 50 simulations as well as for random and
Barab\'asi-Albert scale free models with the same number of nodes and
average degree.  It is clear from such results that the designed
models have topological properties strikingly similar to those of the
respective Barab\'asi-Albert models, except for the hierarchical
common degree, which resulted remarkably distinct, exhibiting a peak
near at the higher hierarchical levels.  Slightly higher values of
clustering coefficient are also observed for the design models.

\section{Scale Free by Dynamics}

\begin{figure}
 \begin{center} 
   \includegraphics[scale=.4]{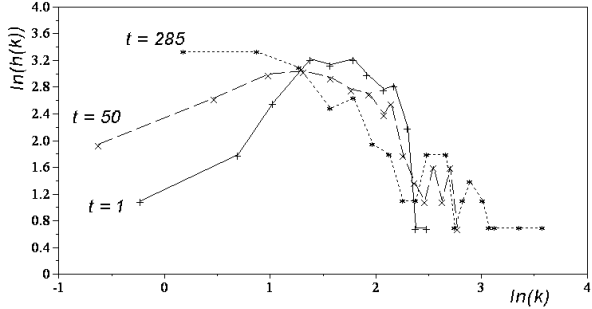}
   \caption{Three stages of the subnetwork degree evolution by using
   the suggested node exchange dynamics. ~\label{fig_dyn}}
 \end{center}
\end{figure}

The concepts and methods described in the previous sections can also
be used to implement a dynamics of node exchange between the
subnetworks in a partitioned system.  Among the several possibilities,
we investigate the scheme starting with a uniform subnetwork partition
of a random network $\Gamma$ (i.e. each community $i$ initially has
$N_i=N/n$ nodes) and involving sequential random selection of a
subnetwork $c_i$, from which a node is randomly selected (uniform
probability) and moved to (possibly) another subnetwork $c_j$ chosen
with probability proportional to its respective degree $k(c_j)$.  It
is suggested that such a dynamical node exchange can be used to model
several real-world phenomena such as the continuous exchange of
individuals between institutions, e.g. music performers moving from an
ensemble to another, animal species changing their environment, and so
on.  Figure~\ref{fig_dyn} shows the log-log plot of the subnetwork
degree distributions for three successive steps --- i.e. $t=1$, $t=50$
and $t=185$ --- along the node exchange iteractions.  It is clearly
perceived that the left-hand side of the log-log distribution tends to
increase as the nodes are redistributed among the subnetworks.

\section{Concluding Remarks}

The concepts of subnetwork degree~\cite{Costa_gener:2004} as well as
the presently introduced notion of subnetwork partitions, have allowed
interesting developments such as the design and evolution of scale
free subnetworks.  The hierarchical characterization of experimental
results of a designed subnetwork partitioned model indicates that such
networks present similar features to equivalent Barab\'asi-Albert
models, except for the hierarchical common degree, which tended to
present a peak at higher hierarchical levels.  Although we have
concentrated attention on scale free degree distribution, the proposed
concepts and methods can be immediately applied to many other
situations including the design of \emph{community} organized networks
with generic degree distribution.  Because for large values of $N$ the
subnetwork degree can be well-approximated by the product between the
number of nodes inside the subnetwork and the average degree of the
underlying random network, the subnetwork degree distribution
ultimately follows the distribution of the number of nodes in the
subnetworks.  As a consequence, geographical networks where nodes are
uniformly distributed along the space and the subnetworks cover areas
which follow a power law will result naturally scale free.  Another
issue deserving further attention is the dynamical redistribution of
nodes among the subnetworks.

\begin{acknowledgments}
Luciano da F. Costa is grateful to FAPESP
(proc. 99/12765-2), CNPq (proc. 308231/03-1) and the Human Frontier
Science Program (RGP39/2002) for financial support.
\end{acknowledgments}

\bibliography{inst}

\begin{thebibliography}{10}
\expandafter\ifx\csname natexlab\endcsname\relax\def\natexlab#1{#1}\fi
\expandafter\ifx\csname bibnamefont\endcsname\relax
  \def\bibnamefont#1{#1}\fi
\expandafter\ifx\csname bibfnamefont\endcsname\relax
  \def\bibfnamefont#1{#1}\fi
\expandafter\ifx\csname citenamefont\endcsname\relax
  \def\citenamefont#1{#1}\fi
\expandafter\ifx\csname url\endcsname\relax
  \def\url#1{\texttt{#1}}\fi
\expandafter\ifx\csname urlprefix\endcsname\relax\def\urlprefix{URL }\fi
\providecommand{\bibinfo}[2]{#2}
\providecommand{\eprint}[2][]{\url{#2}}

\bibitem[{\citenamefont{Flory}(1941)}]{Flory:1941}
\bibinfo{author}{\bibfnamefont{P.~J.} \bibnamefont{Flory}},
  \bibinfo{journal}{J. Am. Chem. Soc.} \textbf{\bibinfo{volume}{63}},
  \bibinfo{pages}{3083} (\bibinfo{year}{1941}).

\bibitem[{\citenamefont{Rapoport}(1957)}]{Rapoport:1957}
\bibinfo{author}{\bibfnamefont{A.}~\bibnamefont{Rapoport}},
  \bibinfo{journal}{Bull. Math. Bioph.} \textbf{\bibinfo{volume}{19}},
  \bibinfo{pages}{257} (\bibinfo{year}{1957}).

\bibitem[{\citenamefont{\"Erdos and R\'enyi}(1959)}]{Erdos:1959}
\bibinfo{author}{\bibfnamefont{P.}~\bibnamefont{\"Erdos}} \bibnamefont{and}
  \bibinfo{author}{\bibfnamefont{A.}~\bibnamefont{R\'enyi}},
  \bibinfo{journal}{Publ. Math.} \textbf{\bibinfo{volume}{6}},
  \bibinfo{pages}{290} (\bibinfo{year}{1959}).

\bibitem[{\citenamefont{Albert and Barab\'asi}(2002)}]{Albert_Barab:2002}
\bibinfo{author}{\bibfnamefont{R.}~\bibnamefont{Albert}} \bibnamefont{and}
  \bibinfo{author}{\bibfnamefont{A.~L.} \bibnamefont{Barab\'asi}},
  \bibinfo{journal}{Rev. Mod. Phys.} \textbf{\bibinfo{volume}{74}},
  \bibinfo{pages}{47} (\bibinfo{year}{2002}).

\bibitem[{\citenamefont{Newman}(2003)}]{Newman:2003}
\bibinfo{author}{\bibfnamefont{M.~E.~J.} \bibnamefont{Newman}},
  \bibinfo{journal}{SIAM Review} \textbf{\bibinfo{volume}{45}},
  \bibinfo{pages}{167} (\bibinfo{year}{2003}),
  \bibinfo{note}{cond-mat/0303516}.

\bibitem[{\citenamefont{da~F.~Costa}(2004{\natexlab{a}})}]{Costa_hub:2004}
\bibinfo{author}{\bibfnamefont{L.}~\bibnamefont{da~F.~Costa}}
  (\bibinfo{year}{2004}{\natexlab{a}}), \bibinfo{note}{cond-mat/0405022}.

\bibitem[{\citenamefont{Bagrow and Bollt}(2004)}]{Bollt:2004}
\bibinfo{author}{\bibfnamefont{J.}~\bibnamefont{Bagrow}} \bibnamefont{and}
  \bibinfo{author}{\bibfnamefont{E.}~\bibnamefont{Bollt}}
  (\bibinfo{year}{2004}), \bibinfo{note}{cond-mat/0412482}.

\bibitem[{\citenamefont{Latapy and Pons}(2004)}]{Latapy:2004}
\bibinfo{author}{\bibfnamefont{M.}~\bibnamefont{Latapy}} \bibnamefont{and}
  \bibinfo{author}{\bibfnamefont{P.}~\bibnamefont{Pons}}
  (\bibinfo{year}{2004}), \bibinfo{note}{cond-mat/0412368}.

\bibitem[{\citenamefont{da~F.~Costa}(2004{\natexlab{b}})}]{Costa_gener:2004}
\bibinfo{author}{\bibfnamefont{L.}~\bibnamefont{da~F.~Costa}}
  (\bibinfo{year}{2004}{\natexlab{b}}), \bibinfo{note}{cond-mat/0408076}.

\bibitem[{\citenamefont{da~F.~Costa}(2004{\natexlab{c}})}]{Costa_hier_char:200%
4}
\bibinfo{author}{\bibfnamefont{L.}~\bibnamefont{da~F.~Costa}}
  (\bibinfo{year}{2004}{\natexlab{c}}), \bibinfo{note}{cond-mat/0412761}.

\end{thebibliography}

\end{document}